\documentstyle[amssymb,aps,prb,preprint]{revtex}

\begin{document}
\draft
\title{Structure factors for the simplest solvable model of polydisperse colloidal
fluids with surface adhesion}
\author{Domenico Gazzillo and Achille Giacometti}
\address{Istituto Nazionale di Fisica della Materia and \\
Facolt\`{a} di Scienze, Universit\`{a} di Venezia, \\
S. Marta DD 2137, I-30123 Venezia, Italy}
\date{\today}
\maketitle

\newpage

\begin{abstract}
Closed analytical expressions for scattering intensity and other global
structure factors are derived for a new solvable model of polydisperse
sticky hard spheres. The starting point is the exact solution of the ``mean
spherical approximation'' for hard core plus Yukawa potentials, in the limit
of infinite amplitude and vanishing range of the attractive tail, with their
product remaining constant. The choice of factorizable coupling (stickiness)
parameters in the Yukawa term yields a simpler ``dyadic structure'' in the
Fourier transform of the Baxter factor correlation function $q_{ij}(r)$,
with a remarkable simplification in all structure functions with respect to
previous works. The effect of size and stickiness polydispersity is analyzed
and numerical results are presented for two particular versions of the
model: i) when all polydisperse particles have a single, size-independent,
stickiness parameter, and ii) when the stickiness parameters are
proportional to the diameters. The existence of two different regimes for
the average structure factor, respectively above and below a generalized
Boyle temperature which depends on size polydispersity, is recognized and
discussed. Because of its analycity and simplicity, the model may be useful
in the interpretation of small-angle scattering experimental data for
polydisperse colloidal fluids of neutral particles with surface adhesion.
\end{abstract}

\newpage

\newpage

\section{INTRODUCTION}

A theoretical determination of scattering intensity and structure factors
for fluids with a large number $p$ of components is, in general, a difficult
task. In particular, this is true when some kind of polydispersity is
present, as occurs in colloidal or micellar solutions \cite
{Pusey91,Lowen94,Nagele96}. Size polidispersity means that macroparticles of
a same chemical species exhibit several different dimensions within a
discrete or continuous set of possible values. Interaction polydispersity
then denotes a similar, and usually correlated, dispersion of parameters
(charges, etc.) defining the strength of interaction potentials. Even when
all macroparticles belong to a unique chemical species, a polydisperse fluid
must therefore be treated as a multi-component mixture, with very large $p$
values (of order $10^1\div 10^3$) or, in the infinite-component limit $%
p\rightarrow \infty ,$ with an idealized continuous distribution of some
properties \cite{Stell79}.

In structural studies on polydisperse colloidal fluids, a key role can be
played by the models for which the Ornstein-Zernike (OZ) integral equations
of the liquid state \cite{Hansen86} admit analytical solutions leading to
closed expressions of scattering functions for any finite $p$ and even for $%
p\rightarrow \infty .$ A sufficient condition for this is that the Fourier
transforms $\widehat{q}_{ij}\left( k\right) $ of the functions $q_{ij}(r)$,
solutions of the Baxter factorized version of the OZ equations \cite
{Baxter70,Hiroike79}, have a peculiar mathematical form, which we refer to
as {\it dyadic} structure \cite{Gazzillo97} and will be illustrated in
Section II C. Using the dyadicity, the explicit inversion of a related $%
p\times p$ matrix $\widehat{{\bf Q}}\left( k\right) $ is always possible for
arbitrary $p$ and closed analytical expressions for the ``partial''
structure factors $S_{ij}(k)$ \cite{Ashcroft67} can be obtained. The
scattering intensity and other ``global'' structure factors are then
calculated as weighted sums of all partial structure factors.

Usually, the above sums are performed numerically by evaluating $p\left(
p+1\right) /2$ independent contributions $S_{ij}(k)$ at each $k$ \cite
{Stell79,Griffith86,Senatore85,Ginoza99}. This procedure becomes numerically
demanding for large $p$, as required in polydisperse mixtures. On the other
hand, we stress that the dyadicity property also enables an alternative
route (followed in the present work) which avoids the explicit computation
of individual $S_{ij}(k)$. The weighted sums can, in fact, be worked out
analytically, by a procedure originally proposed by Vrij \cite{Vrij79} and
referred to as ``Vrij's summation'' hereafter. The resulting closed
analytical expressions of ``global'' scattering functions hold true for {\it %
any} number $p$ of components, can be easily applied to polydisperse fluids
even in the limit $p\rightarrow \infty ,$ and are particularly suitable to
fit experimental scattering data. Vrij \cite{Vrij79} first obtained a closed
expression for the scattering intensity of polydisperse hard spheres (HS)
within the Percus-Yevick (PY) approximation. Gazzillo {\it et al.} \cite
{Gazzillo97} derived similar formulas for polydisperse charged hard spheres
(CHS), by using the corresponding analytical solution within the ``mean
spherical approximation'' (MSA).

In the present paper we extend the approach previously exploited for
polydisperse HS \cite{Vrij79} and CHS \cite{Gazzillo97} to polydisperse
``sticky'' hard spheres (SHS). This simple model adds to an interparticle
hard core repulsion an infinitely strong attraction at contact, and can be
applied to real colloidal fluids of neutral spherical particles with a van
der Waals (or dispersion) force of attraction, working at very short
distances.

Baxter \cite{Baxter68} proposed the one-component original version of this
model, solved the OZ equation with the PY closure and found that such a
system presents a liquid-gas phase transition. The adhesive contribution in
Baxter's Hamiltonian is defined by a particular limiting case of a
square-well tail in which the depth goes to infinity as the width goes to
zero, in such a way that the contribution to the second virial coefficient
remains finite but not zero (Baxter limit). Stell \cite{Stell91} found that
SHS of equal diameter in the Baxter limit, when treated exactly rather than
in the PY approximation, are not thermodynamically stable. Nevertheless, the
PY solution for SHS as a useful colloidal model has received a continuously
growing interest in the last two decades \cite{Stell91,Regnaut89}, partially
because of its capability to exhibit a gas-liquid phase transition.

Perram and Smith \cite{Perram75} and Barboy and Tenne \cite{Barboy79}
extended Baxter's work to $p$-component fluids, using the same kind of
Hamiltonian and the PY approximation. Santos {\it et al.} \cite{Santos98}
developed a rational function approximation to go beyond the PY
approximation and derived improved expressions for the radial distribution
functions and structure factors of SHS mixtures.

Unfortunately, the PY analytical solution for mixtures requires the
determination of a set of unknown {\it density-dependent} parameters $%
\lambda _{ij}$, related, through $p(p+1)/2$ coupled quadratic equations, to
other parameters $\tau _{ij}$ which appear in the potentials as
monotonically increasing functions of the temperature $T$ and whose inverses
measure the degree of adhesion (``stickiness'') of interacting spheres. In
most cases the coefficients $\lambda _{ij}$ for given $\tau _{ij}$ can only
be found numerically, and this feature limits the applicability of the
SHS-PY model to small $p$ values. As a matter of fact the number of actual
applications to polydisperse fluids is very limited. We are aware of a study
by Robertus {\it et al. }\cite{Robertus89} on small angle x-ray scattering
from microemulsions, with polydispersity represented by $p=9$ components, a
work by Penders and Vrij \cite{Penders90} on turbidity of silica particles,
and an investigation by Duits {\it et al. }\cite{Duits91} on small angle
neutron scattering from sterically stabilized silica particles dispersed in
benzene. To simplify the numerical determination of the set $\left\{ \lambda
_{ij}\right\} $, all these papers treat the special case of a single
stickiness parameter, $\tau $, independent of particle size.

In general, the SHS-PY solution for mixtures does not have the dyadic
structure which allows the {\it analytic} inversion of $\widehat{{\bf Q}}%
\left( k\right) $ required to get closed expressions for structure factors
of polydisperse systems$.$ To recover the dyadicity and obtain an explicitly
solvable model, Herrera and Blum \cite{Herrera91} used the {\it ad hoc }%
assumption $\lambda _{ij}=\lambda _i\lambda _j$, in a study on polydisperse
CHS with sticky interactions under a MSA/PY closure.

On the other hand, apart from Baxter's original definition \cite{Baxter68},
there exists a second version of the SHS model, proposed in the
one-component case by Brey {\it et al.} \cite{Brey87} and Mier-y-Teran {\it %
et al.} \cite{Mier89}. Here, the adhesive part of the potential is defined
as the limit of a Yukawa tail when both amplitude and inverse range tend to
infinity, with their ratio remaining constant. The analytic solution was
obtained within the MSA closure \cite{Mier89}, and Ginoza and Yasutomi \cite
{Ginoza96} discussed its relationship to Baxter's PY solution. Recently,
Tutschka and Kahl \cite{Tutschka98} investigated the multi-component version
of this second SHS model and presented MSA expressions of structure
functions for both the discrete (finite $p$) and the continuous ($%
p\rightarrow \infty $) polydisperse case. These authors pointed out that the
dyadic structure of the SHS-MSA solution can be ensured {\it a priori} by
imposing from the outset a Berthelot-type rule \cite{Hansen86} on the
coupling (stickiness) parameters $\gamma _{ij\text{ }}$of the Yukawa tail,
i.e. $\gamma _{ij\text{ }}=(\gamma _{ii\text{ }}\gamma _{jj\text{ }})^{1/2}.$

Within the same framework of Yukawa-MSA models, the present paper has a
threefold aim: i) we shall show that a new choice of Yukawa coupling
parameters, $Y_i$, slightly different from Tutschka and Kahl's ones \cite
{Tutschka98}, can produce an even simpler solvable model of SHS, with a
remarkable simplification of all analytical results; ii) we shall obtain
closed analytical expressions for scattering intensity and other ``global''
structure factors of SHS by extending the formalism successfully employed
for HS and CHS; iii) we shall present numerical applications not only in the
case of equal stickiness for all particles, but also when a simple
size-dependence of the stickiness parameters $Y_i$ is assumed.

The interplay among stickiness attraction, size polydispersity, and hard
core repulsion gives rise to a rather complex behaviour. Nonetheless we
shall present a simple unified description of these results, hinging on the
introduction of a generalized Boyle temperature.

The paper is organized as follows. In the next section the basic theory on
structure factors, integral equations and dyadic matrices will be briefly
recalled. The SHS model, its MSA solution under the assumption of
factorizable coefficients, expressions for scattering intensity and other
``global'' structure factors will be given in Sec. III. Numerical results
are included in Sec. IV, while the last section is devoted to a summary and
some conclusive remarks.

\section{BASIC FORMALISM}

\subsection{Structure factors and scattering intensity}

All scattering functions of multicomponent or polydisperse fluids with
spherically symmetric interparticle potentials can be expressed in terms of
``partial'' structure factors $S_{ij}(k)$, such as the Ashcroft-Langreth
ones \cite{Ashcroft67}

\begin{equation}
S_{ij}\left( k\right) =\delta _{ij}+H_{ij}\left( k\right) =\delta
_{ij}+\left( \rho _i\rho _j\right) ^{1/2}\widetilde{h}_{ij}\left( k\right) .
\end{equation}

\noindent Here, $k$ is the magnitude of the scattering vector, $\delta _{ij}$
the Kronecker delta$,$ $\rho _i$ the number density of species $i,$ $%
\widetilde{h}_{ij}\left( k\right) $ the three-dimensional Fourier transform
of the total correlation function, $h_{ij}\left( r\right) =g_{ij}\left(
r\right) -1,$ with $g_{ij}\left( r\right) $ being the radial distribution
function between two particles of species $i$ and $j$ at distance $r$.

The knowledge of the $S_{ij}\left( k\right) $ allows to calculate the
scattering intensity as well as some ``global'' structure factors. The {\it %
coherent scattering intensity }$I(k)$ for a $p$-component fluid is given by 
\cite{Pusey91,Vrij79}

\begin{equation}
R(k)\equiv I\left( k\right) /V=\rho \sum_{i,j=1}^p\left( x_ix_j\right)
^{1/2}F_i\left( k\right) F_j^{*}\left( k\right) S_{ij}\left( k\right) ,
\end{equation}

\noindent where $V$ is the volume, $\rho =\sum_m\rho _m$ the total number
density, while $x_i=\rho _i/\rho $ and $F_i(k)$ denote the molar fraction
and form factor of species $i,$ respectively (the asterisk means complex
conjugation). The {\it measurable average\ structure factor }is then defined
from the Rayleigh ratio $R(k)$ as \cite{Pusey91,Vrij79}

\begin{equation}
S_{{\rm M}}\left( k\right) =R(k)\ /\left[ \ \rho P(k)\right] .  \label{s3}
\end{equation}

\noindent with $P(k)=\sum_{m=1}^px_m\left| F_m\left( k\right) \right| ^2.$
As a third useful quantity, we consider the Bathia-Tornton {\it %
number-number structure factor} \cite{Bhatia70}, which is related to number
density fluctuations:

\begin{equation}
S_{{\rm NN}}\left( k\right) =\sum_{i,j=1}^p\left( x_ix_j\right)
^{1/2}S_{ij}\left( k\right) .
\end{equation}

\noindent The definition of other global structure factors may be found in
Ref. 3. Clearly, $R(k)$, $S_{{\rm M}}\left( k\right) $ and $S_{{\rm NN}%
}\left( k\right) $ involve a unique kind of weighted sum, i.e.,

\begin{equation}
\sum_{i,j=1}^pw_i\left( k\right) w_j^{*}\left( k\right) S_{ij}\left(
k\right) ,
\end{equation}

\noindent with $w_i\left( k\right) $ being equal to $\rho _i^{1/2}F_i\left(
k\right) ,\left[ x_i/P(k)\right] ^{1/2}F_i\left( k\right) ,$ and $%
x_i{}^{1/2},$ respectively.

\subsection{Integral equations in Baxter form}

Integral equations of the liquid state theory represent a powerful
theoretical tool to get the $h_{ij}\left( r\right) $ required to calculate
the partial structure factors $S_{ij}\left( k\right) $. The Ornstein-Zernike
(OZ) integral equations relate the $h_{ij}\left( r\right) $ functions to the
simpler direct correlation functions $c_{ij}\left( r\right) .$ For fluids
with spherically symmetric interactions, these equations are \cite
{Hansen86,Lee88}

\begin{equation}
h_{ij}\left( r\right) =c_{ij}\left( r\right) +\sum_{m=1}^p\rho _m\int d{\bf r%
}^{\prime }\ c_{im}\left( r^{\prime }\right) h_{mj}\left( |{\bf r-r}^{\prime
}|\right)
\end{equation}

\noindent and can be solved only when coupled with an approximate second
relationship (a ``closure'') among $c_{ij}\left( r\right) $, $h_{ij}\left(
r\right) $ and interparticle potential $u_{ij}\left( r\right) $ \cite
{Hansen86,Lee88}.

By Fourier transformation, the OZ convolution equations become, in $k$-space,

\begin{equation}
\left[ {\bf I}+{\bf H}\left( k\right) \right] \left[ {\bf I}-{\bf C}\left(
k\right) \right] ={\bf I,}
\end{equation}

\noindent where $C_{ij}\left( k\right) \equiv \left( \rho _i\rho _j\right)
^{1/2}\widetilde{c}_{ij}\left( k\right) $ and $\widetilde{c}_{ij}\left(
k\right) $ is the Fourier transform of $c_{ij}\left( r\right) .$ If ${\bf S}%
\left( k\right) $ denotes the symmetric matrix with elements $S_{ij}\left(
k\right) $, then we get

\begin{equation}
{\bf S}\left( k\right) ={\bf I}+{\bf H}\left( k\right) =\left[ {\bf I}-{\bf C%
}\left( k\right) \right] ^{-1},  \label{i3}
\end{equation}

\noindent with ${\bf I}$ being the unit matrix of order $p.$ The $%
S_{ij}\left( k\right) $ can therefore be expressed in terms not only of $%
\widetilde{h}_{ij}\left( k\right) $, but also of $\widetilde{c}_{ij}\left(
k\right) .$ However, in this paper we shall use a third representation of $%
S_{ij}\left( k\right) $ based upon the Baxter factor correlation functions $%
q_{ij}(r)$ \cite{Baxter70}. By means of a Wiener-Hopf factorization of ${\bf %
I}-{\bf C}\left( k\right) ,$ Baxter transformed the OZ equations for HS
fluids into an equivalent, but easier to solve, form \cite{Baxter70}. Later
on these equations were extended by Hiroike to any spherically symmetric
potentials, without using the Wiener-Hopf factorization \cite{Hiroike79}.
Baxter factorization reads

\begin{equation}
{\bf I}-{\bf C}(k)=\widehat{{\bf Q}}^T\left( -k\right) \widehat{{\bf Q}}%
\left( k\right) ,  \label{i4}
\end{equation}

\noindent where $\widehat{{\bf Q}}\left( k\right) $ has the form 
\begin{equation}
\widehat{{\bf Q}}\left( k\right) ={\bf I}-\widetilde{{\bf Q}}\left( k\right)
={\bf I}-\int_{-\infty }^{+\infty }dr\ e^{ikr}{\bf Q}\left( r\right) ,
\end{equation}

\noindent with $Q_{ij}(r)=2\pi \left( \rho _i\rho _j\right) ^{1/2}q_{ij}(r)$
($\ \widehat{{\bf Q}}^T$ is the transpose of $\widehat{{\bf Q}}\ $). Note
that $\widehat{Q}_{ij}\left( -k\right) =\widehat{Q}_{ij}^{*}\left( k\right)
. $ For fluids of particles with spherically symmetric interactions
including HS repulsions (i.e., $u_{ij}\left( r\right) =+\infty $ when $%
r<\sigma _{ij}\equiv (\sigma _i+\sigma _j)/2,$ with $\sigma _i$ = hard
sphere diameter of species $i$), the Baxter equations in $r$-space are

\begin{equation}
\left\{ 
\begin{array}{l}
rc_{ij}\left( |r|\right) =-q_{ij}^{\ \prime }(r)+2\pi \sum_m\rho
_m\int_{L_{mi}}^\infty dt\ q_{mi}\left( t\right) q_{mj}^{\ \prime }\left(
r+t\right) , \\ 
\qquad \qquad \qquad \qquad \qquad \qquad \qquad \\ 
rh_{ij}\left( |r|\right) =-q_{ij}^{\ \prime }(r)+2\pi \sum_m\rho
_m\int_{L_{im}}^\infty dt\ q_{im}\left( t\right) \left( r-t\right) \\ 
\qquad \qquad \qquad \qquad \qquad \qquad \times h_{mj}\left( |r-t|\right) ,
\end{array}
\right.
\end{equation}

\noindent where $r>L_{ij}\equiv (\sigma _i-\sigma _j)/2$ and the prime
denotes differentiation with respect to $r$.

Using Eqs. ($\ref{i3}$) and ($\ref{i4}$), we get ${\bf S}(k)=\widehat{{\bf Q}%
}^{-1}\left( k\right) \left[ \widehat{{\bf Q}}^{-1}\left( -k\right) \right]
^T$ and the partial structure factors can be written as

\begin{equation}
S_{ij}(k)=\sum_m\widehat{Q}_{im}^{-1}\left( k\right) \widehat{Q}%
_{jm}^{-1}\left( -k\right) =\sum_m\widehat{Q}_{im}^{-1}\left( k\right)
\left[ \widehat{Q}_{jm}^{-1}\left( k\right) \right] ^{*}.  \label{i7}
\end{equation}

\noindent On defining

\begin{equation}
s_m(k)=\sum_{i=1}^pw_i(k)\widehat{Q}_{im}^{-1}\left( k\right) ,  \label{i8}
\end{equation}

\noindent all the ``global'' structure functions can then be expressed as 
\begin{equation}
\sum_{i,j=1}^pw_i\left( k\right) w_j^{*}\left( k\right) S_{ij}\left(
k\right) =\sum_{m=1}^ps_m(k)s_m^{*}(k).  \label{i9}
\end{equation}

\subsection{Dyadic matrices and Vrij's summation}

The main problem of these analytical calculations hinges on the inversion of
the matrix $\widehat{{\bf Q}}\left( k\right) ={\bf I}-\widetilde{{\bf Q}}%
\left( k\right) $, which usually becomes a formidable task with increasing
the number $p$ of components. In a particular case, however, the inverse $%
\widehat{{\bf Q}}^{-1}\left( k\right) $ can be easily found for any size of
the original matrix. This occurs when $\widehat{Q}_{ij}(k)$ is a {\it dyadic}
(or Jacobi) matrix, i.e. when it has the peculiar mathematical structure 
\begin{equation}
\widehat{Q}_{ij}=\delta _{ij}+\sum_{\mu =1}^na_i^{(\mu )}b_j^{(\mu )}\qquad
(i,j=1,\ldots ,p)  \label{d1}
\end{equation}

\noindent (the dependence on $k$ was omitted for simplicity). We recall that
a matrix $T_{ij}=a_ib_j$ formed by the direct product of two vectors is
often referred to as a {\it dyad}, ${\bf {ab},}$ while a linear combination
of dyads $\sum_\mu \lambda _\mu {\bf {a}^{\left( \mu \right) }b^{\left( \mu
\right) }}$ is called a {\it dyadic} \cite{Mathews65}. Moreover, we shall
refer to a sum of $n$ dyads as a {\it n-dyadic.}

We caution the reader that, since $\widehat{Q}_{ij}(k)=\delta _{ij}-2\pi
\left( \rho _i\rho _j\right) ^{1/2}\widehat{q}_{ij}\left( k\right) ,$ where $%
\widehat{q}_{ij}\left( k\right) $ is the unidimensional Fourier transform of 
$q_{ij}\left( r\right) ,$ Eq. (\ref{d1}) actually requires that $\widehat{q}%
_{ij}\left( k\right) $ is a $n$-dyadic matrix (of order $p$), but in the
following we shall use the same terminology for $\widehat{Q}_{ij}(k)$ as
well. The dyadicity is actually present in some solvable models of fluid
mixtures: $\widehat{q}_{ij}\left( k\right) $ is $2$-dyadic in the PY
solution for neutral HS\cite{Stell79,Vrij79}, and $3$-dyadic in the MSA
solution for CHS \cite{Blum77}.

The dyadic matrices have some special properties, which have already been
partially discussed in Ref. 8. Here we recall the main points along with new
additional features. Let us associate to the matrix of order $p$ of Eq. (\ref
{d1}), $\widehat{{\bf Q}}{\bf \ =I}+\sum_{\mu =1}^n{\bf a}^{(\mu )}{\bf b}%
^{(\mu )},$ a matrix ${\bf D}_{{\rm Q}}$ of order $n$ with elements 
\begin{equation}
d_{\alpha \beta }=\delta _{\alpha \beta }+{\bf a}^{(\alpha )}\cdot {\bf {b}}%
^{(\beta )}\qquad \left( \alpha ,\beta =1,\ldots ,n\right) ,
\end{equation}
\noindent where the dot denotes the usual scalar product of vectors, i.e. $%
{\bf a}^{(\alpha )}\cdot {\bf {b}}^{(\beta )}{\bf =}\sum_{m=1}^pa_m^{(\alpha
)}b_m^{(\beta )}.$ A first property is that any $n$-dyadic matrix $\widehat{%
{\bf Q}}$ of order $p$ has always rank $n$, irrespective of $p.$ Moreover,
its determinant $\left| \widehat{{\bf Q}}\right| ,$ which is of order $p,$
turns out to be equal to the determinant $D_{{\rm Q}}\equiv \left| {\bf D}_{%
{\rm Q}}\right| $ of order $n$ (with $n\ll p$ in multi-component fluids). A
second property yields the explicit form of the elements of the inverse
matrix $\widehat{{\bf Q}}^{-1},$ as

\begin{equation}
\widehat{Q}_{ij}^{-1}=\delta _{ij}-\frac 1{D_{{\rm Q}}}\sum_{\alpha
=1}^n\sum_{\beta =1}^na_i^{(\alpha )}b_j^{(\beta )}\left| {\bf D}_{{\rm Q}%
}\right| ^{\alpha \beta },  \label{d2}
\end{equation}

\noindent where $\left| {\bf D}_{{\rm Q}}\right| ^{\alpha \beta }$ is the
cofactor of the ($\alpha ,\beta $)th element in ${\bf D}_{{\rm Q}}$ and the
double sum contains $n^2$ determinants of order $n$. Clearly, this
expression could also be rewritten as a sum of only $n$ determinants, i.e., $%
\widehat{Q}_{ij}^{-1}=\delta _{ij}-\sum_{\alpha =1}^na_i^{(\alpha )}\widehat{%
D}_j^{(\alpha )}/D_{{\rm Q}},$ where $\widehat{D}_j^{(\alpha )}\equiv
\sum_{\beta =1}^nb_j^{(\beta )}\left| {\bf D}_{{\rm Q}}\right| ^{\alpha
\beta }$ is the determinant obtained from $D_{{\rm Q}}$ by replacing the $%
\alpha $-th {\it row} with $b_j^{(1)},...,b_j^{(n)}$ . Alternatively, one
could write $\widehat{Q}_{ij}^{-1}=\delta _{ij}-\sum_{\beta =1}^nD_i^{(\beta
)}b_j^{(\beta )}/D_{{\rm Q}},$ with $D_i^{(\beta )}\equiv \sum_{\alpha
=1}^na_i^{(\alpha )}\left| {\bf D}_{{\rm Q}}\right| ^{\alpha \beta }$ being
obtained from $D_{{\rm Q}}$ by replacing the $\beta $-th {\it column} with $%
a_i^{(1)},...,a_i^{(n)}$ (Cramer rule).

All above expressions reflect the remarkable fact that $\widehat{{\bf Q}}%
^{-1}$ is a $n$-dyadic matrix as well, and it is indeed this property
enabling a successful outcome of Vrij's summation for ``global'' structure
functions. Our starting point is a reformulation of Eq. (\ref{d2}) in terms
of a determinant of order $n+1:$

\begin{equation}
\widehat{Q}_{ij}^{-1}=\frac 1{D_{{\rm Q}}}\left| 
\begin{array}{lllll}
\delta _{ij}\quad & \ \qquad b_j^{(1)} & \qquad b_j^{(2)} & \cdots & \qquad
b_j^{(n)} \\ 
a_i^{(1)}\quad & ~1+{\bf a}^{(1)}\cdot {\bf {b}}^{(1)} & \qquad {\bf a}%
^{(1)}\cdot {\bf {b}}^{(2)} & \cdots & \qquad {\bf a}^{(1)}\cdot {\bf {b}}%
^{(n)} \\ 
a_i^{(2)} & \qquad {\bf a}^{(2)}\cdot {\bf {b}}^{(1)}{\bf \quad } & \ 1+{\bf %
a}^{(2)}\cdot {\bf {b}}^{(2)} & \cdots & \qquad {\bf a}^{(2)}\cdot {\bf {b}}%
^{(n)} \\ 
\ \vdots & \ \qquad \quad \vdots & \ \qquad \quad \vdots & ~\vdots & \
\qquad \quad \vdots \\ 
a_i^{(n)} & \qquad {\bf a}^{(n)}\cdot {\bf {b}}^{(1)} & \qquad {\bf a}%
^{(n)}\cdot {\bf {b}}^{(2)} & \cdots & 1+{\bf a}^{(n)}\cdot {\bf {b}}^{(n)}
\end{array}
\right| ,  \label{d3}
\end{equation}

\noindent where ${\bf D}_{{\rm Q}}$ is included as a submatrix. From Eqs. (%
\ref{i8}) and (\ref{d3}), one then gets

\begin{equation}
s_m=\frac 1{D_{{\rm Q}}}\left| 
\begin{array}{lllll}
\quad w_m & b_m^{(1)} & b_m^{(2)} & \cdots & b_m^{(n)} \\ 
{\bf w}\cdot {\bf {a}}^{(1)}\quad & d_{11} & d_{12} & \cdots & d_{1n} \\ 
{\bf w}\cdot {\bf {a}}^{(2)} & d_{21}{\bf \quad } & d_{22} & \cdots & d_{2n}
\\ 
\ \quad \vdots & ~\vdots & ~\vdots & ~\ \vdots & ~\vdots \\ 
{\bf w}\cdot {\bf {a}}^{(n)} & d_{n1} & d_{n2} & \ \cdots & d_{nn}
\end{array}
\right| .  \label{d4}
\end{equation}

\noindent To perform the sum over $m$ required in Eq. (\ref{i9}), we expand
this determinant along the first row to get

\begin{equation}
s_m=w_m+\sum_{\alpha =1}^nb_m^{(\alpha )}C_\alpha ,  \label{d5}
\end{equation}

\noindent where $C_\alpha \equiv T_\alpha /D_{{\rm Q}}$ and $T_\alpha $ $%
\left( \alpha =0,1,\ldots ,n\right) $ is the cofactor of the element $\left(
1,\alpha +1\right) $th in the determinant of Eq. (\ref{d4}). Clearly, $%
T_0=D_{{\rm Q}}$. Using Eqs. (\ref{i9}) and (\ref{d5}), the searched final
result is

\begin{equation}
\sum_{i,j=1}^pw_iw_j^{*}S_{ij}={\bf w}\cdot {\bf {w}}^{*}+2\sum_{\alpha =1}^n%
\mathop{\rm Re}%
\ \left[ {\bf w}^{*}\cdot {\bf {b}}^{(\alpha )}C_\alpha \right]
+\sum_{\alpha =1}^n\sum_{\beta =1}^n{\bf b}^{(\alpha )}\cdot {\bf {b}}%
^{(\beta )*}\ C_\alpha C_\beta ^{*},  \label{d6}
\end{equation}

\noindent where $%
\mathop{\rm Re}%
\left[ ...\right] $ denotes the real part of a complex number. Vrij \cite
{Vrij79} first performed a similar computation and derived a closed
expression for the scattering intensity of HS mixtures. Our expression
generalizes Vrij's one as well as that of Ref. 8 for the scattering
intensity of CHS. It is simpler and more compact and can be used to
calculate any ``global'' structure function, for mixtures with {\it any}
number $p$ of components. It involves a sum of only $(n+1)^2$ terms, and
depends on $p$ through some averages represented by scalar products of
vectors. Only the number of terms contained in these averages increases with
increasing $p,$ and hence the application to polydisperse mixtures is
straightforward.

\section{THE MODEL}

\subsection{Sticky hard spheres as a limit of Yukawa particles}

A fluid of SHS can be derived from particles interacting via HS plus Yukawa
(HSY) attractive potentials, i.e.

\begin{equation}
-\beta u_{ij}(r)=\left\{ 
\begin{array}{lll}
+\infty , &  & 0<r<\sigma _{ij} \\ 
\beta A_{ij}e^{-\mu (r-\sigma _{ij})}/r, &  & r>\sigma _{ij}
\end{array}
,\right. \qquad  \label{y1}
\end{equation}

\noindent in the limit $\mu \rightarrow +\infty $ with $\beta A_{ij}=\mu
K_{ij}$ and $K_{ij}$ independent of $\mu $. Here, $\beta =(k_BT)^{-1},$
where $k_B$ is the Boltzmann constant, and all $A_{ij}$ are positive, with $%
A_{ji}=A_{ij}$ and $K_{ji}=K_{ij}$, as required by the symmetry condition $%
u_{ji}(r)=u_{ij}(r)$. This approach is convenient, since the Baxter
equations for HSY mixtures have been solved analytically \cite{Blum78}, for
any finite $\mu ,$ within the MSA closure, which adds to the exact hard core
condition, $h_{ij}\left( r\right) =-1$ for $r<\sigma _{ij},$ the approximate
relationship $c_{ij}\left( r\right) =-\beta u_{ij}(r)$ for $r>\sigma _{ij}.$
The solution is

\begin{equation}
q_{ij}(r)=\left\{ 
\begin{array}{l}
0,\qquad \qquad r<L_{ij} \\ 
\frac 12a_i(r-\sigma _{ij})^2+(b_i+a_i\sigma _{ij})(r-\sigma
_{ij})+C_{ij}\left[ e^{-\mu (r-\sigma _{ij})}-1\right] \\ 
\qquad \qquad \qquad +D_{ij}e^{-\mu (r-\sigma _{ij})},\qquad L_{ij}<r<\sigma
_{ij} \\ 
D_{ij}e^{-\mu (r-\sigma _{ij})},\qquad r>\sigma _{ij}
\end{array}
\right.
\end{equation}

\noindent where the coefficients are determined by a complicate set of
equations \cite{Blum78}. From these, however, it can be shown that, as $\mu
\rightarrow +\infty ,$ then $C_{ij}\rightarrow -D_{ij}$ and $%
D_{ij}\rightarrow \beta A_{ij}/\mu =K_{ij}.$ The MSA solution for SHS is
therefore

\begin{equation}
q_{ij}(r)=\left\{ 
\begin{array}{l}
\frac 12a_i(r-\sigma _{ij})^2+(b_i+a_i\sigma _{ij})(r-\sigma
_{ij})+K_{ij},\qquad L_{ij}\leq r\leq \sigma _{ij} \\ 
0,\qquad \qquad \qquad \qquad \text{elsewhere}
\end{array}
\right.  \label{y3}
\end{equation}

\begin{equation}
a_i=\frac 1\Delta +\frac{3\xi _2\sigma _i}{\Delta ^2}-\frac{12\zeta _i}\Delta
,\qquad b_i=\left( \frac 1\Delta -a_i\right) \frac{\sigma _i}2,
\end{equation}

\begin{equation}
\xi _m=\frac \pi 6\sum_{i=1}^p\rho _i\sigma _i^m,\qquad \zeta _i=\frac \pi 6%
\sum_{j=1}^p\rho _j\sigma _jK_{ij}\ ,\qquad \Delta =1-\xi _3.
\end{equation}

To calculate $\widehat{Q}_{ij}^{-1}(k),$ we need the unidimensional Fourier
transform $\widehat{q}_{ij}\left( k\right) $, i.e.

\begin{eqnarray}
\widehat{q}_{ij}\left( k\right) &=&-e^{{\rm i}X_i}\ \left\{ \left( 1+\frac{%
3\xi _2\sigma _i}\Delta -12\zeta _i\right) \frac 1{4\Delta }\sigma _j^3\ 
\frac{j_1(X_j)}{X_j}\right.  \label{y6} \\
&&\left. +\frac 1{4\Delta }\sigma _i\sigma _j^2\left[ \ j_0(X_j)-{\rm i}%
j_1(X_j)\right] -K_{ij}\sigma _j\ j_0(X_j)\right\} ,  \nonumber
\end{eqnarray}

\noindent where $X_m\equiv k\sigma _m/2,$ $\ j_0(x)=\sin x/x$ and $\
j_1(x)=\left( \sin x-x\cos x\right) /x^2$ are spherical Bessel functions,
and ${\rm i}$ - when it is not a subscript - is the imaginary unit.

\subsection{Factorizable coefficients}

In general $\widehat{q}_{ij}\left( k\right) ,$ as defined in Eq. (\ref{y6}),
does not have the required dyadic structure, due to the presence of $K_{ij}$
in the last term. To overcome this difficulty, Tutschka and Kahl \cite
{Tutschka98} proposed the following {\it Ansatz:}

\begin{equation}
K_{ij}=\gamma _{ij}\sigma _{ij}^2,\qquad \text{with\qquad }\gamma _{ij\text{ 
}}=(\gamma _{ii\text{ }}\gamma _{jj\text{ }})^{1/2}\quad \text{%
(Berthelot-rule),}  \label{f1}
\end{equation}

\noindent which yields $K_{ij}=\gamma _{ii}^{1/2}\gamma _{jj}^{1/2}(\sigma
_i^2+2\sigma _i\sigma _j+\sigma _j^2)/4.$ Consequently, the last term of Eq.
(\ref{y6}) splits into three independent contributions, and $\widehat{q}%
_{ij}\left( k\right) $ turns out to be 5-dyadic, in spite of the fact that
in the HS limit (no adhesion) it is only 2-dyadic.

We first note that a great simplification occurs with the factorization

\begin{equation}
K_{ij}=Y_iY_j  \label{f2}
\end{equation}

\noindent (all $Y_m\geq 0$). In this case, the last term of Eq. (\ref{y6})
generates only one contribution, and $\widehat{q}_{ij}\left( k\right) $
becomes simply 3-dyadic and, in a particular case to be discussed later on,
even 2-dyadic. The $K_{ij}$ defined by Eq. (\ref{f2}) satisfy the
Berthelot-rule, i.e. $K_{ij}=(K_{ii}K_{jj})^{1/2}.$ Note that the stickiness
parameters $\gamma _{mm\text{ }}$are dimensionless \cite{Tutschka98}, while
the $Y_m$ are lengths.

Factorizable adhesive parameters have already been considered by Yasutomi
and Ginoza \cite{Yasutomi96} and Herrera {\it et al. }\cite{Herrera98} in
studies on adhesive-HSY fluids, although no expressions for structure
functions were given. Since in those papers $K_{ij}=GG_iG_j$, the
relationship with our coefficients is simply given by $Y_m=G^{1/2}G_m.$

Using Eq. (\ref{f2}), we get $\zeta _i=\xi _2^Y\ Y_i\ $ with $\xi _2^Y\equiv
\left( \pi /6\right) \sum_{j=1}^p\rho _j\sigma _jY_j$ (dimensionally
analogous to $\xi _2$), and therefore

\begin{eqnarray}
\widehat{Q}_{ij}(k) &=&\delta _{ij}-2\pi \left( \rho _i\rho _j\right) ^{1/2}%
\widehat{q}_{ij}\left( k\right)  \nonumber \\
&=&\delta _{ij}+\rho _i^{1/2}e^{{\rm i}X_i}\ \left\{ \left( 1+\frac{3\xi
_2\sigma _i}\Delta -{\rm i}\frac{k\sigma _i}2-12\xi _2^Y\ Y_i\ \right) \frac %
\pi {2\Delta }\sigma _j^3\ \frac{j_1(X_j)}{X_j}\right.  \label{f3} \\
&&\left. +\frac \pi {2\Delta }\sigma _i\sigma _j^2\ j_0(X_j)-2\pi
Y_iY_j\sigma _j\ j_0(X_j)\right\} \rho _j^{1/2},  \nonumber
\end{eqnarray}

\noindent which has the required dyadic structure $\widehat{Q}_{ij}=\delta
_{ij}+\sum_{\mu =1}^na_i^{(\mu )}b_j^{(\mu )}.$ We emphasize that the
decomposition into $a_i^{(\mu )}$ and $b_j^{(\mu )}$ is not unique. Our
choice allows an easy comparison with the corresponding results for
polydisperse HS \cite{Vrij79} and CHS \cite{Gazzillo97}. After defining

\begin{eqnarray}
\alpha _m(k) &=&\frac \pi {2\Delta }\sigma _m^3\frac{j_1(X_m)}{X_m}, 
\nonumber \\
\beta _m^{(0)}(k) &=&\frac \pi {2\Delta }\sigma _m^2j_0(X_m),\qquad \beta
_m(k)=\beta _m^{(0)}(k)+\left( \frac{3\xi _2}\Delta -{\rm i}\frac k2\right)
\alpha _m(k),  \label{f4} \\
\delta _m^{(0)}(k) &=&-2\pi \sigma _mY_m\ j_0(X_m),\qquad \delta
_m(k)=\delta _m^{(0)}(k)-12\xi _2^Y\ \alpha _m(k),  \nonumber
\end{eqnarray}

\noindent $\widehat{Q}_{ij}(k)$ may be rewritten as

\begin{equation}
\widehat{Q}_{ij}(k)=\delta _{ij}+\rho _i^{1/2}e^{{\rm i}X_i}\left[ \alpha
_j(k)+\sigma _i\beta _j(k)+Y_i\delta _j(k)\right] \rho _j^{1/2},  \label{f5}
\end{equation}

\noindent and the corresponding decomposition is

\begin{equation}
\begin{array}{lll}
a_i^{(1)}=\rho _i^{1/2}e^{{\rm i}X_i}, & \qquad & b_j^{(1)}=\rho
_j^{1/2}\alpha _j, \\ 
a_i^{(2)}=\rho _i^{1/2}e^{{\rm i}X_i}\sigma _{i\ }, & \qquad & 
b_j^{(2)}=\rho _j^{1/2}\beta _j, \\ 
a_i^{(3)}=\rho _i^{1/2}e^{{\rm i}X_i}Y_i\ , & \qquad & b_j^{(3)}=\rho
_j^{1/2}\delta _j.
\end{array}
\label{f6}
\end{equation}

If we are interested in the scattering intensity, then Eq. (\ref{d4}) with $%
w_m=\rho _m^{1/2}F_m$ yields

\begin{equation}
s_m=\frac{\rho _m^{1/2}}{D_{{\rm Q}}}\left| 
\begin{array}{llll}
\ F_m & \qquad \ \alpha _m & \ \qquad \beta _m & \qquad \ \delta _m \\ 
\left\{ F\right\} & \quad 1+\left\{ \alpha \right\} & \qquad \left\{ \beta
\right\} & \qquad \left\{ \delta \right\} \\ 
\left\{ \sigma F\right\} & \qquad \left\{ \sigma \alpha \right\} & \quad
1+\left\{ \sigma \beta \right\} & \qquad \left\{ \sigma \delta \right\} \\ 
\left\{ YF\right\} & \qquad \left\{ Y\alpha \right\} & \qquad \left\{ Y\beta
\right\} & \quad 1+\left\{ Y\delta \right\}
\end{array}
\right|  \label{f7}
\end{equation}

\noindent where $D_{{\rm Q}}$ coincides with the cofactor of the (1,1)th
element in the $4\times 4$ determinant, and

\begin{equation}
~\left\{ fg\right\} \equiv \sum_{m=1}^p\rho _me^{{\rm i}X_m}f_mg_m=\rho
\left\langle e^{{\rm i}X}fg\right\rangle ,\qquad ~\left\langle
f\right\rangle \equiv \sum_{m=1}^px_mf_m.
\end{equation}

\noindent Here and in the following, angular brackets $\left\langle \cdots
\right\rangle $ denote compositional averages over the distribution of
particles (this notation differs from that of Ref. 8, which also contains
misprints corrected in Refs. 35,36).

With the chosen decomposition of $\widehat{Q}_{ij}(k)$ the stickiness
contributions are confined in the last row and the last column of the
determinant of Eq. (\ref{f7}). When adhesion is turned off, all these
elements vanish apart from their diagonal term, and the $4\times 4$
determinant essentially reduces to the $3\times 3$ HS one. For numerical
computation, it is convenient, following Ref. 8, to simplify $s_m$ by using
elementary transformations which do not alter the value of the determinant.
If we add to the third column the second one multiplied by $-\left( 3\xi
_2/\Delta -{\rm i}k/2\right) $ and to the fourth column the second one
multiplied by $12\xi _2^Y$, then $s_m$ becomes

\begin{equation}
s_m=\frac{\rho _m^{1/2}}{D_{{\rm Q}}}\left| 
\begin{array}{llll}
\ F_m & \qquad \ \alpha _m & \ \qquad \beta _m^{(0)} & \qquad \ \delta
_m^{(0)} \\ 
\left\{ F\right\} & \quad 1+\left\{ \alpha \right\} & \qquad \left\{ \beta
^{(0)}\right\} -3\xi _2/\Delta +{\rm i}k/2 & \qquad \left\{ \delta
^{(0)}\right\} +12\xi _2^Y \\ 
\left\{ \sigma F\right\} & \qquad \left\{ \sigma \alpha \right\} & \quad
1+\left\{ \sigma \beta ^{(0)}\right\} & \qquad \left\{ \sigma \delta
^{(0)}\right\} \\ 
\left\{ YF\right\} & \qquad \left\{ Y\alpha \right\} & \qquad \left\{ Y\beta
^{(0)}\right\} & \quad 1+\left\{ Y\delta ^{(0)}\right\}
\end{array}
\right| ,  \label{f9}
\end{equation}

\noindent where $D_{{\rm Q}}$ has been changed accordingly. Expanding the $%
4\times 4$ determinant along the first line and inserting it into Eq. (\ref
{i9}), the final result for the scattering intensity of SHS is

\begin{eqnarray}
R(k)/\rho &=&\left\langle F^2\right\rangle +\left\langle \alpha
^2\right\rangle \left| C_1\right| ^2+\left\langle \beta ^{(0)2}\right\rangle
\left| C_2\right| ^2+\left\langle \delta ^{(0)2}\right\rangle \left|
C_3\right| ^2  \nonumber \\
&&+2%
\mathop{\rm Re}%
\left[ \left\langle F\alpha \right\rangle C_1+\left\langle F\beta
^{(0)}\right\rangle C_2+\left\langle F\delta ^{(0)}\right\rangle C_3\right.
\label{f10} \\
&&\left. +\ \left\langle \alpha \beta ^{(0)}\right\rangle
C_1C_2^{*}+\left\langle \alpha \delta ^{(0)}\right\rangle
C_1C_3^{*}+\left\langle \beta ^{(0)}\delta ^{(0)}\right\rangle
C_2C_3^{*}\right] ,  \nonumber
\end{eqnarray}

\noindent where form factors have been assumed to be real quantities, as is
indeed the case for spherical homogeneous scattering cores. On the r.h.s.
all $k$ arguments have been omitted for simplicity, and the $C_\nu (k)$ have
already been defined with reference to Eq. (\ref{d5}).

The expression for the average structure factor $S_{{\rm M}}\left( k\right) $
is then obtained after division of the Rayleigh ratio $R(k)$ by $\rho
P(k)=\rho \left\langle F^2(k)\right\rangle .$ Moreover, the Bathia-Thornton
number-number structure factor $S_{{\rm NN}}\left( k\right) $ can easily be
derived by setting all $F_m=1$ everywhere into the expression of $R(k)/\rho $%
, with the result

\begin{eqnarray}
S_{{\rm NN}}\left( k\right) &=&1+\rho \left\{ \left\langle \alpha
^2\right\rangle \left| {\cal C}_1\right| ^2+\left\langle \beta
^{(0)2}\right\rangle \left| {\cal C}_2\right| ^2+\left\langle \delta
^{(0)2}\right\rangle \left| {\cal C}_3\right| ^2\right.  \nonumber \\
&&+2%
\mathop{\rm Re}%
\left[ \left\langle \alpha \right\rangle {\cal C}_1+\left\langle \beta
^{(0)}\right\rangle {\cal C}_2+\left\langle \delta ^{(0)}\right\rangle {\cal %
C}_3\right.  \label{f11} \\
&&\left. \left. +\ \left\langle \alpha \beta ^{(0)}\right\rangle {\cal C}_1%
{\cal C}_2^{*}+\left\langle \alpha \delta ^{(0)}\right\rangle {\cal C}_1%
{\cal C}_3^{*}+\left\langle \beta ^{(0)}\delta ^{(0)}\right\rangle {\cal C}_2%
{\cal C}_3^{*}\right] \right\} ,  \nonumber
\end{eqnarray}

\noindent where the ${\cal C}_\nu $ are the analogues of the $C_\nu $
appearing in $R(k)$.

\section{RESULTS FOR POLYDISPERSE FLUIDS}

\subsection{Size distribution}

For SHS fluids containing only one chemical species, size polydispersity
simply means the presence of a multiplicity of possible diameters. In a
``discrete'' representation of polydispersity the number $p$ of different
diameters is very large but finite, and $x_{i\text{ }}$is the fraction $%
N_i/N $ of particles having diameter $\sigma _i$. On the other hand, a
theoretical representation with infinitely many components ($p\rightarrow
\infty )$ and ``continuously'' distributed diameters is also possible and
often used.

Although all formulas of previous Sections refer to a finite number $p$ of
components, the polydisperse continuous limit of such expressions can
immediately be inferred by the {\it replacement rules} $x_\alpha \rightarrow 
{\rm d}x=f(\sigma ){\rm d}\sigma $ and $\sum_\alpha x_\alpha ...\rightarrow
\int {\rm d}\sigma f(\sigma )...$, where $f(\sigma ){\rm d}\sigma $ is the
fraction $dN/N$ of particles with diameter in the interval $\left( \sigma
,\sigma +{\rm d}\sigma \right) $. As {\it molar fraction density function} $%
f(\sigma )$ we choose the Schulz distribution \cite{Beurten81,DAguanno91}

\begin{equation}
f(\sigma )=\frac a{\Gamma (a)\left\langle \sigma \right\rangle }\left( a\ 
\frac \sigma {\left\langle \sigma \right\rangle }\right) ^{a-1}\exp \left(
-a\ \frac \sigma {\left\langle \sigma \right\rangle }\right) \;\;\quad (a>1),
\end{equation}

\noindent
where $\Gamma $ is the gamma function\cite{Abramowitz72}, $\left\langle
\sigma \right\rangle $ the average diameter, $a=1/s_\sigma ^2,$ and $%
s_\sigma =\left[ \left\langle \sigma ^2\right\rangle -\left\langle \sigma
\right\rangle ^2\right] ^{1/2}/\left\langle \sigma \right\rangle $ measures
the degree of size polydispersity. In the monodisperse limit, $s_\sigma =0,$
the distribution becomes a Dirac delta function centered at $\left\langle
\sigma \right\rangle $. The Schulz function allows an easy analytic
evaluation of some averages $\int {\rm d}\sigma f(\sigma )...$, such as the
moments $\left\langle \sigma ^m\right\rangle $, which obey a simple relation
for $m\geq 1$, i.e. 
\begin{equation}
\left\langle \sigma ^m\right\rangle =\left[ 1+(m-1)s_\sigma ^2\right]
\left\langle \sigma \right\rangle \left\langle \sigma ^{m-1}\right\rangle
=\left\langle \sigma \right\rangle ^m\prod_{j=1}^{m-1}M_j\ ,  \label{sd2}
\end{equation}

\noindent with $M_j\equiv 1+js_\sigma ^2\ .$

In most cases, however, analytical integration is hardly feasible, and
numerical integration brings back to discrete expressions with large $p$, of
order $10^2-10^3$. In practice, the ``discrete'' representation of
polydispersity is the most convenient for numerical purposes, and all
formulas of the previous Sections can be employed by assuming $x_\alpha
=f(\sigma _\alpha )\Delta \sigma $, where $\Delta \sigma $ is the grid size
of numerical integration.

For fluids with Schulz-distributed diameters the packing fraction, $\eta
\equiv \xi _3=(\pi /6)\rho \left\langle \sigma ^3\right\rangle $, can be
written as $\eta =\eta _{{\rm mono}}\left( 1+s_\sigma ^2\right) \left(
1+2s_\sigma ^2\right) ,$ with $\eta _{{\rm mono}}=(\pi /6)\rho \left\langle
\sigma \right\rangle ^3$.

\subsection{Stickiness distribution}

On a dimensional basis, the parameters $Y_i$ must be lengths. Moreover, $%
K_{ij}=Y_iY_j$ must be proportional to $\beta =(k_BT)^{-1}$. If we assume,
for simplicity, that stickiness polydispersity and size polydispersity are
fully correlated, then the most natural choice for $Y_i$ is

\begin{equation}
Y_i=\gamma _{0\ }\sigma _i\ ,  \label{sp1}
\end{equation}

\noindent with the dimensionless proportionality factor

\begin{equation}
\gamma _0\ =\left( \frac{\varepsilon _0}{k_BT}\right) ^{1/2}=\frac{%
Y_{\left\langle \sigma \right\rangle }}{\left\langle \sigma \right\rangle },
\label{sp2}
\end{equation}

\noindent where $\varepsilon _0$ denotes an energy and $Y_{\left\langle
\sigma \right\rangle }$ is the stickiness parameter of particles with
diameter $\left\langle \sigma \right\rangle .$ This implies that

\begin{equation}
K_{ij}=\gamma _0^2\ \sigma _i\sigma _j=\frac{\varepsilon _0}{k_BT}\ \sigma
_i\sigma _j\ =\frac 1{T^{*}}\ \sigma _i\sigma _j\ ,\   \label{sp3}
\end{equation}

\noindent where we have also introduced a reduced temperature $T^{*}=\left(
k_BT/\varepsilon _0\right) \equiv 1/\gamma _0^2$.

The model of Eq. (\ref{sp1}) will be compared with the one of SHS
polydisperse in size but not in stickiness (on the analogy of Refs. 20 and
21). In this simpler case all particles have the same $Y_i=Y_{\left\langle
\sigma \right\rangle }=\gamma _0\left\langle \sigma \right\rangle ,$ and the
degree of stickiness polydispersity $s_Y$, defined similarly to $s_\sigma ,$
vanishes.

Both these models may be regarded as particular cases (for $\alpha =0$ and $%
\alpha =1$) of a more general size-dependence given by

\begin{equation}
Y_i=Y_{\left\langle \sigma \right\rangle }\left( \frac{\sigma _i}{%
\left\langle \sigma \right\rangle }\right) ^\alpha =\gamma _0\ \frac{\sigma
_i^\alpha }{\left\langle \sigma \right\rangle ^{\alpha -1}},  \label{sp4}
\end{equation}

\noindent with $\alpha \geq 0$. We have examined this generalization for $%
\alpha =2$ and $\alpha =3$, but for the purposes of the present paper we
restrict our analysis only to the cases $\alpha =0$ and $\alpha =1.$

The choice $Y_i=\gamma _0\ \sigma _i$ has very interesting properties.
First, the corresponding distribution of $Y$-values, related to the size
distribution $f_\sigma $ as $f_Y\equiv dN/dY=f_\sigma d\sigma /dY$, is a
Schulz function as well, with $\left\langle Y\right\rangle =Y_{\left\langle
\sigma \right\rangle }$ and $s_Y=s_\sigma $. A second more important fact is
that only in this special case $\widehat{Q}_{ij}(k)$, in general 3-dyadic
for SHS, becomes simply 2-dyadic, i.e.

\begin{equation}
\widehat{Q}_{ij}(k)=\delta _{ij}+\rho _i^{1/2}e^{{\rm i}X_i}\ \left\{
A_i(k)\alpha _j(k)+G_0\sigma _i\beta _j^{(0)}(k)\right\} \rho _j^{1/2},
\label{sp5}
\end{equation}

\noindent with

\[
A_i(k)=1+\left( \frac{3\xi _2G_0}\Delta -{\rm i}\frac k2\right) \sigma _i\ , 
\]
\begin{equation}
G_0\equiv 1-4\gamma _0^2\Delta =1-\frac{4\varepsilon _0}{k_BT}\left[ 1-\eta
_{{\rm mono}}\left( 1+s_\sigma ^2\right) \left( 1+2s_\sigma ^2\right) \right]
\label{sp6}
\end{equation}

\noindent (now $\xi _2^Y=\gamma _0\ \xi _2$). It is remarkable that this
expression for $\widehat{Q}_{ij}(k)$ differs from the HS one only for the
presence of $G_0$ ($G_0=1$ for HS). Now the natural dyadic decomposition
becomes

\begin{equation}
\begin{array}{lll}
a_i^{(1)}=\rho _i^{1/2}e^{{\rm i}X_i}A_i, &  & b_j^{(1)}=\rho _j^{1/2}\alpha
_j, \\ 
a_i^{(2)}=\rho _i^{1/2}e^{{\rm i}X_i}G_0\sigma _{i\ }, &  & b_j^{(2)}=\rho
_j^{1/2}\beta _j^{(0)},
\end{array}
\end{equation}

\noindent while the $4\times 4$ determinant appearing in Eq. (\ref{f7})
reduces to a $3\times 3$ one, with a consequent simplification of the
formulas for $R(k)/\rho $, $S_{{\rm M}}\left( k\right) $ and $S_{{\rm NN}%
}\left( k\right) $ (all terms depending on subscript 3 vanish in Eqs. (\ref
{f10}) and (\ref{f11})).

\subsection{Numerical results}

Because of its importance in the analysis of experimental scattering data,
we have focused on the measurable average structure factor $S_{{\rm M}%
}\left( k\right) $. The scattering cores inside the particles have been
assumed to be spherical and homogeneous, with form factors $F_m=\ V_m^{{\rm %
scatt}}\Delta n_m3j_1(X_m^{{\rm scatt}})/X_m^{{\rm scatt}},$ where $X_m^{%
{\rm scatt}}=k\sigma _m^{{\rm scatt}}/2,$ $\sigma _m^{{\rm scatt}}\leq
\sigma _m$ is the diameter of a scattering core of species $m$, $V_m^{{\rm %
scatt}}=\left( \pi /6\right) \left( \sigma _m^{{\rm scatt}}\right) ^3$ its
volume, and $\Delta n_m\ $its scattering contrast with respect to the
solvent. For mixtures with several components belonging to only one chemical
species, as in the present paper, $\Delta n_m$ is the same for all
particles. For simplicity, we have taken $\sigma _m^{{\rm scatt}}=\sigma _m$.

The polydisperse SHS model depends on the following parameters: the packing
fraction $\eta $, the strength $\gamma _0$ of the adhesive interaction, the
average diameter $\left\langle \sigma \right\rangle ,$ and the two degrees
of polydispersity $s_\sigma $ and $s_Y$. In all numerical calculations we
have adopted dimensionless variables, with lengths expressed in units of $%
\left\langle \sigma \right\rangle .$ To understand the influence of each
parameter on $S_{{\rm M}}(k)$, it is instructive to first recall the
behaviour of a sequence of simpler systems, starting from monodisperse hard
spheres and adding in the first two cases either surface attraction or size
polydispersity.

\subsubsection{Monodisperse HS and SHS}

In pure fluids all particles are equal ($s_\sigma =0=s_Y$), $\eta =\eta _{%
{\rm mono}}$, and $S_{{\rm M}}\left( k\right) =S_{{\rm mono}}\left( k\right) 
$ with no form factor involved. Figures 1 and 2 depict the dependence of $S_{%
{\rm mono}}\left( k\right) $ on the parameters ($\eta ,\gamma _0$). Fig. 1
illustrates the evolution of $S_{{\rm mono}}\left( k\right) ,$ as $\eta $
increases from low values up to the freezing one, in the well known case of
monodisperse HS of diameter $\sigma $ without stickiness ($\gamma _0=0$).
Here we have exploited the PY solution \cite{Hansen86}, which, for HS,
coincides with the MSA one. In Fig. 2 the dependence of $S_{{\rm mono}%
}\left( k\right) $ on $\eta $ is displayed for monodisperse SHS, at two
fixed $\gamma _0$ values, i.e. 0.5 and 0.7, corresponding to $\gamma
_0^2=\varepsilon _0/\left( k_BT\right) =0.25$ and $0.49$, or to $T^{*}=4$
and $2.\,04$ ($\gamma _0$ and $T^{*}$ have been defined in Eqs. (\ref{sp2})
and (\ref{sp3}), respectively).

In all these cases (Figs. 1, 2a and 2b), as $\eta $ increases at fixed $%
\gamma _0$, the first peak height and amplitudes of all subsequent
oscillations increase, but the behaviour near the origin depends on $\gamma
_0$, as will be discussed in more detail shortly.

On the other hand, the effect of increasing $\gamma _0$ (i.e. increasing the
adhesive attraction or decreasing $T$) at fixed $\eta $ can be seen by
comparing, for instance, the solid curves ($\eta =0.2$) of Figs. 1 and 2. As 
$\gamma _0$ increases, the first peak and subsequent maxima are shifted to
larger $k$ values, and their amplitudes change as well. However, the most
significant effect on $S_{{\rm mono}}\left( k\right) $ occurs near the
origin. Here, $S_{{\rm mono}}\left( 0\right) $ substantially increases and
becomes the global maximum at large $\gamma _0$. This behaviour can be
understood from the explicit expression of $S_{{\rm mono}}\left( 0\right) $,
which reads

\begin{eqnarray}
S_{{\rm mono}}\left( 0\right) &=&\left[ \widehat{Q}_{{\rm mono}}\left(
0\right) \right] ^{-2}=\frac{\left( 1-\eta \right) ^4}{\left[ 1+2\eta
-12\gamma _0^2\ \eta \left( 1-\eta \right) \right] ^2}  \label{r1} \\
&=&K_T/K_T^{{\rm id}}=\rho k_BTK_T.  \nonumber
\end{eqnarray}

\noindent Since $S_{{\rm mono}}\left( 0\right) $ is related to the
isothermal osmotic compressibility $K_T$ and to the density fluctuations 
\cite{March76}, its drastic increase signals the approach to a gas-liquid
phase transition. The critical point can be obtained from the spinodal line,
defined by $S_{{\rm mono}}^{-1}\left( 0\right) =0$, and the critical
parameters turn out to be: $\eta _{{\rm c}}=\left( \sqrt{3}-1\right)
/2\simeq 0.37$ and $\gamma _{0{\rm c}}^2=\left( \sqrt{3}+2\right) /6\simeq
0.62$ \cite{Brey87} (corresponding to $\gamma _{0{\rm c}}\simeq 0.79$, or to
the reduced {\it critical temperature} $T_{{\rm c}}^{*}\simeq 1.61$).

The combined influence of $\eta $ and $\gamma _0$ can be observed going back
to Fig. 2. On defining the {\it Boyle temperature} $T_{{\rm B}}^{*}$ as the
one where the attractive and repulsive forces balance each other in such a
way that the second virial coefficient $B_2$ vanishes \cite{Penders90}, we
note that the temperatures corresponding to $\gamma _0=0.5$ and $0.7$ lie,
respectively, above and below $T_{{\rm B}}^{*}$ (but in both cases above $T_{%
{\rm c}}^{*}$). In fact, for this monodisperse model it is easy to see, from
the low-density expansion of $S_{{\rm mono}}\left( 0\right) =1+\rho 
\widetilde{h}\left( 0\right) \simeq 1-2B_2\rho +{\cal O}(\rho ^2),$ that $%
B_2=4V_{{\rm HS}}(1-3\gamma _0^2)=4V_{{\rm HS}}\left( 1-3/T^{*}\right) $
with $V_{{\rm HS}}=(\pi /6)\sigma ^3$, and therefore $T_{{\rm B}}^{*}=3$ ($%
\gamma _{0{\rm B}}^2=1/3$ or $\gamma _{0{\rm B}}\simeq 0.58$). Fig. 2
suggests the existence of two different ``regimes'' for $S_{{\rm mono}%
}\left( k\right) $ above and below the Boyle temperature, respectively. When 
$T^{*}>T_{{\rm B}}^{*}$ or, equivalently, $\gamma _0<\gamma _{0{\rm B}}$
(``weak-attraction regime'', as in Fig. 2a) the fluid behaves like pure HS
without stickiness. Here repulsive forces are dominant, $B_2>0$, and
compressibility and density fluctuations, along with the whole $S_{{\rm mono}%
}\left( k\right) $ near the origin, decrease with increasing $\eta $. When $%
T_{{\rm c}}^{*}<T^{*}<T_{{\rm B}}^{*}$ or, equivalently, $\gamma _{0{\rm B}%
}<\gamma _0<\gamma _{0{\rm c}}$ (``strong-attraction regime''), one finds $%
B_2<0,$ while the balance between attractive and repulsive forces becomes
more complex. In this case $S_{{\rm mono}}\left( k\right) $ near the origin
has a non-monotonic dependence on $\eta $, as in Fig. 2b. Here,
compressibility and density fluctuations first increase with $\eta $, in
agreement with the low-density expansion of $S_{{\rm mono}}\left( 0\right) $%
. Then, an inversion occurs at $\eta _0=(6-2T^{*})/(6+T^{*})$ ($\simeq 0.24$
when $T^{*}=2.04$) and afterwards $S_{{\rm mono}}\left( 0\right) $
decreases. In other words, below $T_{{\rm B}}^{*}$ attractive forces seem to
be dominant at low packing fraction, whereas repulsion again prevails at
higher $\eta .$

\subsubsection{Polydisperse HS without stickiness}

Fig. 3 refers to polydisperse HS without surface adhesion ($\gamma _0=0$).
Size polydispersities $s_\sigma =0.1,0.3$ have been employed here and in the
following, since values in this range are rather common in experimental data
from colloidal fluids. The two Schulz distributions have been discretized
with a grid size $\Delta \sigma /\left\langle \sigma \right\rangle =0.02,$
and truncated where $f(\sigma )\Delta \sigma \approx 10^{-8}$, i.e. at $%
\sigma _{{\rm cut}}/\left\langle \sigma \right\rangle =1.68$ and $3.48$,
respectively. Since each diameter characterizes a different component, these
discrete polydisperse mixtures involve $p=85$ and $175$ components. Note
that these numbers of components are much larger than those used with the
SHS-PY model of Ref. 20.

The effect of size polydispersity is considerable \cite{Beurten81}, as
appears from a comparison among Figs. 1, 3a and 3b: with increasing $%
s_\sigma $ at fixed $\eta $, $S_{{\rm M}}\left( k\right) $ slightly
increases in the low-$k $ region, its first peak is reduced and shifted to
smaller $k$ values, and all subsequent oscillations are progressively
dumped, as a result of destructive interference among the several length
scales involved.

\subsubsection{SHS polydisperse in size but not in stickiness}

At this point we study SHS fluids polydisperse in size but monodisperse in
stickiness, with all particles having $Y_i=Y_{\left\langle \sigma
\right\rangle }=\gamma _0\left\langle \sigma \right\rangle $ ($s_\sigma \neq
0$, $\alpha =0\Rightarrow s_Y=0$). This choice will be referred to as Model
I and has been prompted by the SHS-PY investigations of Refs. 20-22, where a
single, size-independent, stickiness parameter was considered.

Figures 4 and 5 illustrate what happens when a surface adhesion (with $%
\gamma _0=0$.$5,0.7$) is added to size polydispersity. Comparison with Fig.
3 ($\gamma _0=0$) shows that the attractive interaction, in the presence of
size polydispersity, produces a further lowering of oscillation amplitudes
in the first peak region and beyond. When $\gamma _0=0.7$ and $s_\sigma =0.3$
(Fig. 5b) all curves exhibit an almost complete flattening in the same range.

Near the origin (for $k\left\langle \sigma \right\rangle \lesssim 5$), for
both considered cases with $\gamma _0=0.5$ (weak-attraction), only a small
increase in $S_{{\rm M}}\left( k\right) $ is found with respect to
polydisperse HS without stickiness (Fig. 3), the relative ordering of all
curves is unchanged and also coincides with that of the corresponding
monodisperse SHS (Fig. 2a). On the other hand, when $\gamma _0=0.7$ and $%
s_\sigma =0.1$ (strong-attraction and low size polydispersity, Fig. 5a) the
behaviour of $S_{{\rm M}}\left( k\right) $ close to the origin strongly
differs from that of polydisperse HS without stickiness and is similar to
the monodisperse SHS case of Fig. 2b. Surface adhesion produces large $S_{%
{\rm M}}\left( 0\right) $ values, which are, however, smaller than the
corresponding monodisperse ones. This means that, when $\gamma _0\neq 0$,
size polydispersity reduces $S_{{\rm M}}\left( k\right) $ even near the
origin (whereas, when $\gamma _0=0$, increasing $s_\sigma $ at fixed $\eta $
determines an increase of $S_{{\rm M}}\left( 0\right) $). This effect of
size polydispersity, in the presence of attraction, is amplified when $%
\gamma _0=0.7$ and $s_\sigma =0.3$ (strong-attraction and high size
polydispersity, Fig. 5b). Now one observes an interesting return to a
``HS-like ordering'' of the curves in the low-$k$ region, as in the case $%
\gamma _0=$ 0.5. This behaviour is peculiar of Model I and will be absent in
Model II to be presented in the next subsection.

\subsubsection{SHS polydisperse both in size and in stickiness}

Next we consider the case of stickiness correlated to the size, according to
the linear law $Y_i=\gamma _{0\ }\sigma _i$ ($\alpha =1,$ $s_Y=s_\sigma \neq
0$). This will be referred to as Model II.

The results for $\gamma _0=$ 0.5 , shown in Fig. 6, are qualitatively
similar to those of Model I (Fig. 4). When $s_Y=s_\sigma =0.1$ the
quantitative differences are very small. However, when $s_Y=s_\sigma =0.3$
the $S_{{\rm M}}(0)$ values lie more clearly above those of Fig. 4b.

For $\gamma _0=$ 0.7 (Fig. 7) the behaviour of $S_{{\rm M}}\left( k\right) $
in the first peak region and beyond is essentially unchanged with respect to
Model I, but near the origin differences are larger and significant. Here,
when $s_Y=s_\sigma =0.1$ the $S_{{\rm M}}\left( k\right) $ curves are
similar to those of Fig. 5a, with larger $S_{{\rm M}}\left( 0\right) $
values (very close to the corresponding monodisperse ones of Fig. 2b), but
as $s_Y=s_\sigma =0.3$ there is a qualitative as well as quantitative
difference with respect to Model I (Fig. 5b). Indeed in the low-$k$ region
the $S_{{\rm M}}\left( k\right) $ curves of Fig. 7b exhibit the same
relative ordering present in the previous case with lower polydispersity
(Fig. 7a) as well as in the corresponding fully monodisperse fluid (Fig.
2b). This persistence in a ``strong-attraction regime'' even at high size
polydispersity constitutes the main difference between Model I and II. Such
a feature can be probably related to the fact that the stickiness
distribution of Model II is skewed towards large $Y_{i\text{ }}$ values
completely absent in Model I, and this asymmetry implies, on average,
stronger attractive forces.

Unfortunately, the behaviour of $S_{{\rm M}}(0)$ in polydisperse models does
not admit any simple thermodynamical interpretation. For mixtures, in fact,
the average structure factor $S_{{\rm M}}\left( k\right) $ depends on the
form factors and $S_{{\rm M}}\left( 0\right) $ is no longer the normalized
compressibility \cite{Nagele96}. Nevertheless, we have been able to account
for the aforesaid difference of ``regimes'' between Model I and II when $%
\gamma _0=0.7$ and $s_\sigma =0.3$ in terms of a single parameter, which
generalizes the Boyle temperature of the monodisperse SHS case.

\subsubsection{Generalized Boyle temperature}

The Boyle temperature of these polydisperse models can be found by deriving
their second virial coefficient $B_2$ from the low-density expansion of $S_{%
{\rm NN}}\left( 0\right) =1-2B_2\rho +{\cal O}(\rho ^2).$ Likewise, to
interpret the behaviour of $S_{{\rm M}}\left( k\right) $ previously
discussed, we start from the low-density expansion of $S_{{\rm M}}\left(
0\right) .$ A straightforward calculation, employing the dyadic formalism of
Sections II and III and not reported here, yields

\begin{equation}
\widehat{Q}_{ij}^{-1}(0)=\delta _{ij}-\rho (x_ix_j)^{1/2}\left[ \frac \pi 6%
\left( \sigma _j^3+3\sigma _i\sigma _j^2\right) -2\pi Y_iY_j\sigma _j\right]
+{\cal O}(\rho ^2),
\end{equation}

\noindent and therefore, from Eq. (\ref{i7}),

\begin{equation}
S_{ij}(0)=\delta _{ij}-\rho (x_ix_j)^{1/2}\left[ \frac \pi 6\left( \sigma
_i+\sigma _j\right) ^3-2\pi \left( Y_iY_j\sigma _j+Y_jY_i\sigma _i\right)
\right] +{\cal O}(\rho ^2).
\end{equation}

\noindent Inserting this result into $S_{{\rm M}}\left( 0\right)
=\sum_{i,j}(x_ix_j)^{1/2}\left[ F_i\left( 0\right) F_j^{*}\left( 0\right)
/P(0)\right] \ S_{ij}\left( 0\right) ,$ and using the above mentioned
expression for $F_m\left( k\right) $, we obtain $S_{{\rm M}}\left( 0\right)
=1-2B_{2,{\rm F}}\rho +{\cal O}(\rho ^2)$, with

\begin{equation}
B_{2,{\rm F}}=\frac \pi 6\frac 1{\left\langle \sigma ^6\right\rangle }\left[
\left\langle \sigma ^6\right\rangle \left\langle \sigma ^3\right\rangle
+3\left\langle \sigma ^5\right\rangle \left\langle \sigma ^4\right\rangle
-12\ \left\langle \sigma ^3Y\right\rangle \left\langle \sigma
^4Y\right\rangle \right] ,  \label{r4}
\end{equation}

\noindent which is the analogue of the second virial coefficient, including
all form factors. The sign of $B_{2,{\rm F}},$ and therefore the behaviour
of $S_{{\rm M}}\left( 0\right) $ at low density (as well as the overall
``regime'' of $S_{{\rm M}}\left( k\right) ),$ depends on $T^{*}$, which is
hidden in the $Y$ terms. On defining a {\it generalized Boyle temperature} $%
T_{{\rm B,F}}^{*}$ as the one where $B_{2,{\rm F}}$ vanishes, and employing
our assumption $Y_i=\gamma _0\sigma _i^\alpha /\ \left\langle \sigma
\right\rangle ^{\alpha -1}$, we find

\begin{equation}
T_{{\rm B,F}}^{*}=\ \frac{12\left\langle \sigma ^{3+\alpha }\right\rangle \
\left\langle \sigma ^{4+\alpha }\right\rangle }{\left[ \left\langle \sigma
^6\right\rangle \left\langle \sigma ^3\right\rangle +3\left\langle \sigma
^5\right\rangle \left\langle \sigma ^4\right\rangle \right] \ \left\langle
\sigma \right\rangle ^{2\left( \alpha -1\right) }}\ .
\end{equation}

\noindent Exploiting Eq. (\ref{sd2}), it follows that

\begin{equation}
\begin{array}{lll}
T_{{\rm B,F}}^{*}=12/\left[ M_4\left( M_5+3M_3\right) \right] &  & \text{for
Model I,} \\ 
T_{{\rm B,F}}^{*}=12M_3/\left( M_5+3M_3\right) &  & \text{for Model II,}
\end{array}
\label{r6}
\end{equation}

\noindent where $M_j$ has been defined with reference to Eq. (\ref{sd2}).
The role of $T_{{\rm B,F}}^{*}$ for $S_{{\rm M}}\left( k\right) $ is the
same as that of $T_{{\rm B}}^{*}$ for $S_{{\rm mono}}\left( k\right) $ of
monodisperse SHS: above $T_{{\rm B,F}}^{*}$ the behaviour is ``HS-like'',
whereas a ``strong-attraction regime'' is found when $T_{{\rm c}%
}^{*}<T^{*}<T_{{\rm B,F}}^{*}$.

Eqs. (\ref{r6}) imply that, in both models, $T_{{\rm B,F}}^{*}$ depends on
the degree of size polydispersity $s_\sigma $. However, whereas $T_{{\rm B,F}%
}^{*}(s_\sigma )$ of Model I is a rapidly decreasing function, in Model II
it exhibits only a very slow decrease asymptotically approaching $18/7\simeq
2.\,57$. Such a difference explains the behaviours displayed in Fig. 5 and
7, which refer to $T^{*}=2.\,04$ (i.e. $\gamma _0=0.7$). In Model I, when $%
s_\sigma =0.1$ one has $T^{*}<T_{{\rm B,F}}^{*}=2.79$, whereas when $%
s_\sigma =0.3$ one finds $T^{*}>T_{{\rm B,F}}^{*}=1.68$. On the other hand,
in both cases the temperature of Model II is below the generalized Boyle
one, the values of $T_{{\rm B,F}}^{*}$ now being $2.99$ and $2.90$ for $%
s_\sigma =0.1$ and $0.3$, respectively.

Finally, it is instructive to compare $T_{{\rm B,F}}^{*}$ with the true
Boyle temperature $T_{{\rm B}}^{*}$ of these polydisperse models$.$ The
second virial coefficient, obtained from the low-density expansion of $S_{%
{\rm NN}}\left( 0\right) ,$ turns out to be $B_2=(\pi /6)\left[ \left\langle
\sigma ^3\right\rangle +3\left\langle \sigma ^2\right\rangle \left\langle
\sigma \right\rangle -12\ \left\langle Y\right\rangle \left\langle \sigma
Y\right\rangle \right] ,$ and one obtains $T_{{\rm B}}^{*}=12/\left[
M_1\left( M_2+3\right) \right] $ and $12/\left( M_2+3\right) $ for Model I
and II, respectively. Note that for these polydisperse fluids it is always $%
T_{{\rm B}}^{*}<\left( T_{{\rm B}}^{*}\right) _{{\rm mono}}=3$, $T_{{\rm B,F}%
}^{*}<T_{{\rm B}}^{*}$ for Model I, and $T_{{\rm B,F}}^{*}>T_{{\rm B}}^{*}$
for Model II, while in the limit of monodisperse fluids $T_{{\rm B,F}%
}^{*}\rightarrow \left( T_{{\rm B}}^{*}\right) _{{\rm mono}}.$

\section{SUMMARY\ AND\ CONCLUSIONS}

In this paper we have presented a new analytically solvable model for
multi-component SHS fluids within the MSA closure, using a hard-core-Yukawa
potential with factorizable coupling parameters (in the appropriate infinite
amplitude and zero range limit). The model is simpler than previous ones
available in the literature, since $\widehat{q}_{ij}(k)$ is in general
3-dyadic (Tutschka and Kahl's model \cite{Tutschka98} was 5-dyadic), with a
consequent great simplification of all analytical formulas.

We have stressed the importance of the ``dyadic structure'' of $\widehat{q}%
_{ij}(k)$ and recalled the properties of matrices with dyadic elements. Such
a feature allows the analytic inversion of $\widehat{Q}_{ij}(k)$ required to
get the partial structure factors $S_{ij}(k)$. Through Vrij's summation,
expressions have then been obtained for global structure functions, such as $%
R(k)$, $S_{{\rm M}}\left( k\right) $ and $S_{{\rm NN}}\left( k\right) $.
These closed analytical formulas, just as their counterparts for
polydisperse HS \cite{Vrij79} and CHS \cite{Gazzillo97}, allow to ``bypass''
the computation of the individual $p(p+1)/2$ partial structure factors,
which may be a rather difficult task for polydisperse systems with large
number of components. Because of their simplicity, our expressions may
therefore represent a very useful tool to fit experimental scattering data
of real colloidal fluids.

While the presented 3-dyadic expressions hold true for any choice of
stickiness parameters $Y_i$, two particular versions of the model have been
analyzed numerically. The first one assumes size polydispersity, but a
single stickiness parameter for all particles (Model I), while the second
one proposes stickiness parameters dependent on the diameters according to a
linear law (Model II). Model I is similar to the SHS-PY models for
polydisperse colloids known in the literature \cite
{Robertus89,Penders90,Duits91}, while Model II is the simplest choice for
size-dependent stickiness parameters.

The combined influence of hard core repulsion, adhesive attraction and
polydispersity can generate a variety of behaviours at the level of
measurable average structure factor $S_{{\rm M}}\left( k\right) $. We have
recognized the existence of two different ``regimes'' for $S_{{\rm M}}\left(
k\right) $ both in monodisperse and in polydisperse SHS fluids. Above a
temperature $T_{{\rm B,F}}^{*}$, which in the monodisperse case coincides
with the Boyle temperature, we have identified a ``weak-attraction''
behaviour, resembling the HS one. In the range below $T_{{\rm B,F}}^{*}$ but
still above the critical temperature, a ``strong-attraction'' regime sets
in, and we have described its features in detail. It is found that $T_{{\rm %
B,F}}^{*}$ is a decreasing function of the degree of size polydispersity $%
s_\sigma .$ It is also worth noting that the behaviour of our SHS-MSA models
in the ``strong-attraction regime'' is in qualitative agreement with that of
the SHS-PY model displayed in Fig.s 3 and 4 of Ref. 20, where the existence
of two different ``regimes'' for $S_{{\rm M}}\left( k\right) $ was, however,
not recognized.

All our numerical results show that size polydispersity strongly affects the
behaviour of $S_{{\rm M}}\left( k\right) $ in the first peak region and
beyond, where the influence of stickiness polydispersity is less
significant. Models I and II are nearly equivalent in this interval of $k$%
-values, whereas they may substantially differ near the origin. The present
study shows that the use of a single stickiness parameter, instead of more
realistic size-dependent ones, may lead to marked differences in the small
angle scattering region at sufficiently high $\gamma _0$, i.e. when
attraction is strong or temperature is low.

In the small $k$ region the adhesive forces and the specific relationship
between stickiness and size parameters have far reaching consequences.
Although very little is known experimentally about the correlation between
stickiness and size, it is reasonable to expect that larger particles
attract each other more strongly. The linear dependence $Y_i=\gamma _{0\
}\sigma _i$ represents the simplest non-trivial choice, but other
possibilities could also be taken into account \cite{Tutschka98}. Duits {\it %
et al.} \cite{Duits91} found that in some cases the SHS-PY model with a
single stickiness parameter, independent of particle size, was unable to fit
their experimental scattering data, and these authors already emphasized the
possible role of stickiness polydispersity as a cause for the observed
deviations. Similar discrepancies between experimental and model $S_{{\rm M}%
}\left( k\right) $ values could be a crucial test for the soundness of our
Model II with respect to Model I, as well as of any other choice for the
stickiness-size functional relationship.

It would be also instructive to compare our SHS-MSA model with other recent
theoretical approaches to polydisperse colloidal fluids. As an example, we
mention the ``optimized random phase approximation'' joined with orthogonal
polynomial expansions, proposed by Lado and coworkers \cite{Leroch99}.

However, the most important advantage of the present model lies in its
simplicity. In particular, version II has special formal properties,
since - only in this case - the expression of $\widehat{q}_{ij}(k)$ becomes
2-dyadic. Therefore version II can indeed be reckoned as the simplest
solvable model for polydisperse SHS and it could be a good candidate to
tackle the issue of thermodynamics and phase stability of these fluids from
a fully analytical point of view. We expect that compact expressions for
pressure, chemical potentials, partial structure factors at $k=0$, as well
as other quantities required to investigate - for instance - sedimentation 
\cite{Bolhuis93}, vapor-liquid equilibrium and demixing in the presence of
polydispersity, can easily be obtained. We hope to accomplish this task in a
forthcoming paper.

\vskip 1cm

\acknowledgments
Partial financial support by the Italian MURST (Ministero
dell'Universit\`{a} e della Ricerca Scientifica through the INFM (Istituto
Nazionale di Fisica della Materia) is gratefully acknowledged.

\begin{figure}[tbp]
\caption{PY ($\equiv$ MSA) structure factor $S_{{\rm mono}}$ of monodisperse
hard spheres without stickiness, plotted as a function of the dimensionless
variable $k \sigma$ at various packing fractions $\eta$. Note that this and
all following figures have the same scale.}
\label{f:Fig1}
\end{figure}

\begin{figure}[tbp]
\caption{MSA structure factor $S_{{\rm mono}}$ of monodisperse sticky hard
spheres, plotted as a function of $k \sigma$ at various packing fractions $%
\eta$, for two values of adhesive strength parameter $\gamma_{0}$ (here, and
in all following figures, the curves of Part b) are shifted upwards by 2.5
units).}
\label{f:Fig2}
\end{figure}

\begin{figure}[tbp]
\caption{PY ($\equiv$ MSA) structure factor $S_{{\rm M}}$ of polydisperse
hard spheres without stickiness, plotted as a function of $k \left\langle
\sigma \right\rangle$ at various packing fractions $\eta$, for two different
degrees of size polydispersity $s_{\sigma}$.}
\label{f:Fig3}
\end{figure}

\begin{figure}[tbp]
\caption{MSA structure factor $S_{{\rm M}}$ of sticky hard spheres
polydisperse in size but monodisperse in stickiness (Model I), plotted as a
function of $k \left\langle \sigma \right\rangle$ at various packing
fractions $\eta$, for $\gamma_{0} = 0.5$ and two different degrees of size
polydispersity $s_{\sigma}$.}
\label{f:Fig4}
\end{figure}

\begin{figure}[tbp]
\caption{As Fig. 4 but for $\gamma_{0} = 0.7$.}
\label{f:Fig5}
\end{figure}

\begin{figure}[tbp]
\caption{MSA structure factor $S_{{\rm M}}$ of sticky hard spheres
polydisperse in size and stickiness (Model II), plotted as a function of $k
\left\langle \sigma \right\rangle$ at various packing fractions $\eta$, for $%
\gamma_{0} = 0.5$ and two different degrees of size polydispersity $%
s_{\sigma}$.}
\label{f:Fig6}
\end{figure}

\begin{figure}[tbp]
\caption{As Fig. 6 but for $\gamma_{0} = 0.7$.}
\label{f:Fig7}
\end{figure}


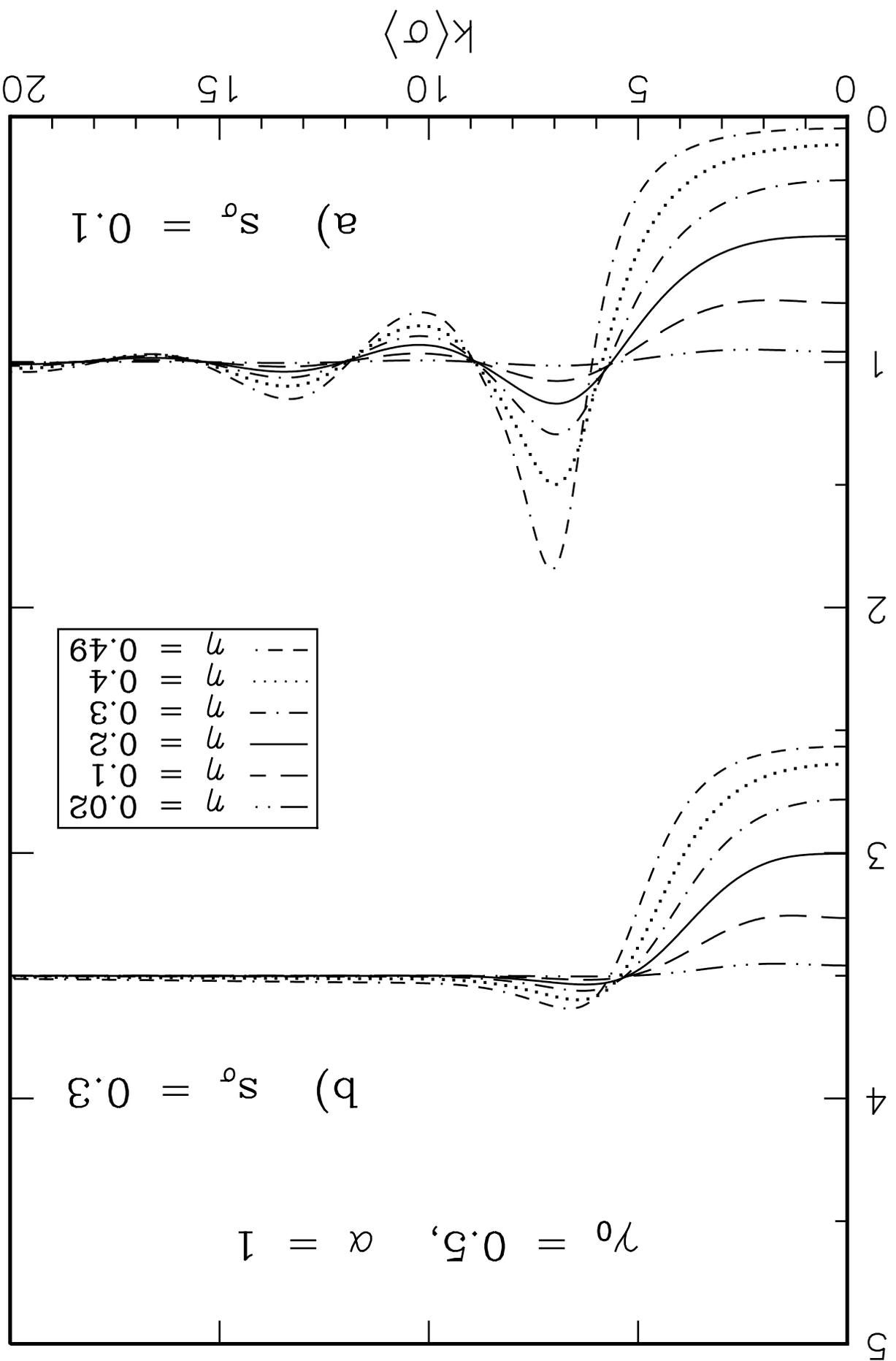
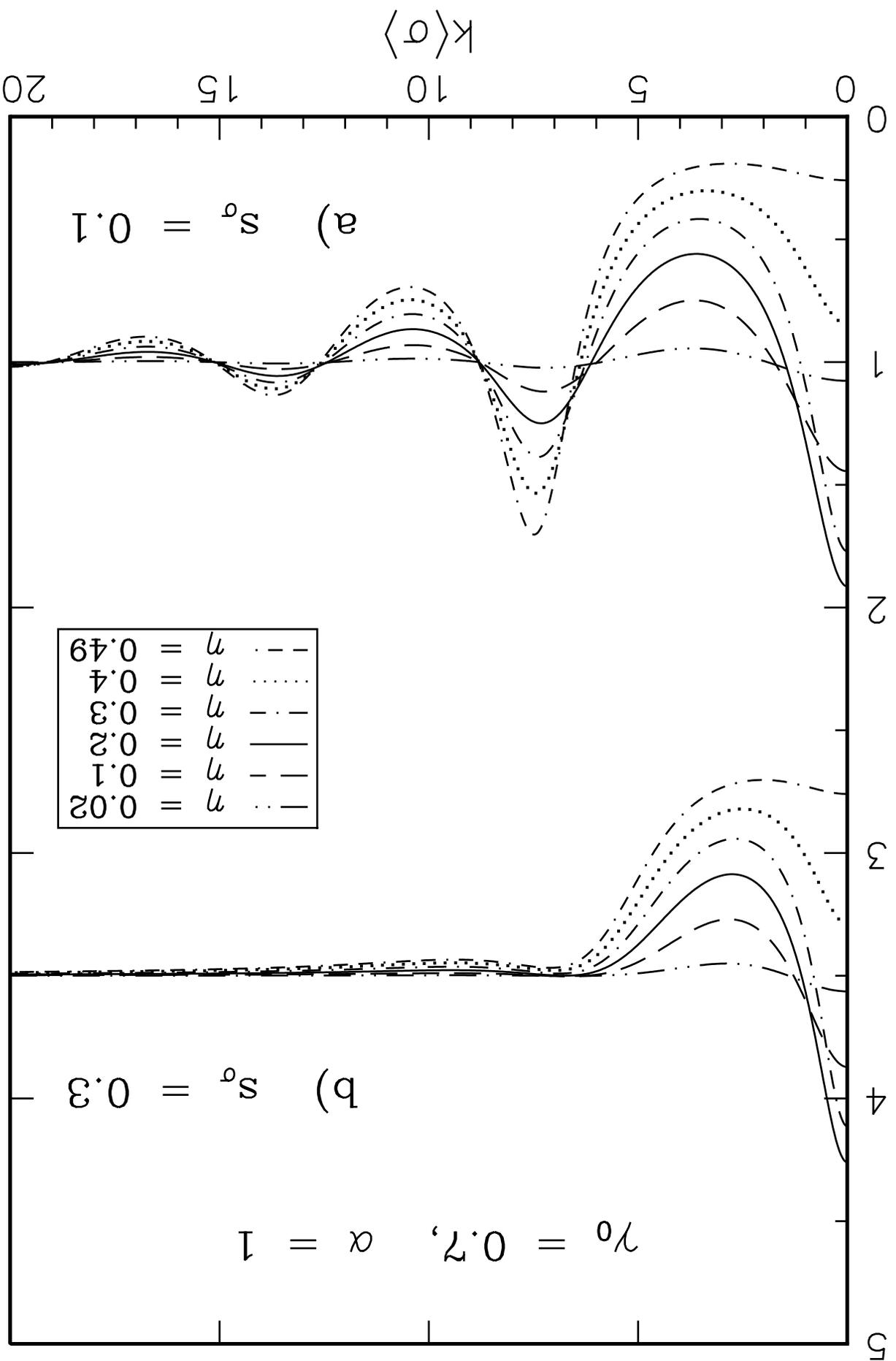
\end{document}